%% file: staerk25d_Tcc-electrode.tex
\documentclass[twocolumn,aps,superscriptaddress,longbibliography]{revtex4}
\usepackage{graphicx}
\usepackage[fleqn]{amsmath}
\usepackage{amssymb}
\usepackage[dvipsnames]{xcolor}
\usepackage[normalem]{ulem}
\definecolor{goodgreen}{rgb}{0.1,0.5,0}
\definecolor{goodred}{rgb}{0.7,0,0}
\usepackage{float}
\usepackage{multirow}
\usepackage{xr}

\usepackage{listings} 

\usepackage{array,mathtools,amssymb,booktabs}
\newcolumntype{C}{>{$}c<{$}}

\AtBeginDocument{%
  \heavyrulewidth=.08em
  \lightrulewidth=.05em
  \cmidrulewidth=.03em
  \belowrulesep=.65ex
  \belowbottomsep=0pt
  \aboverulesep=.4ex
  \abovetopsep=0pt
  \cmidrulesep=\doublerulesep
  \cmidrulekern=.5em
  \defaultaddspace=.5em
}

\usepackage[colorlinks,urlcolor=goodgreen,citecolor=blue,linkcolor=goodred]{hyperref}
\usepackage{bm}
\allowdisplaybreaks
\usepackage{todonotes} 
\usepackage{cleveref} 
\usepackage{siunitx} 
\usepackage{physics} 
\usepackage{csquotes} 

\usepackage{import}
\usepackage{xifthen}
\usepackage{pdfpages}
\usepackage{transparent}

\renewcommand{\vec}[1]{\bm{#1}}
\newcommand{\ljepsilon}{\epsilon}
\renewcommand{\epsilon}{\varepsilon}

\DeclareSIUnit{\calorie}{cal}

\newcommand{\beginsupplement}{%
        \setcounter{table}{0}
        \renewcommand{\thetable}{S\arabic{table}}%
        \setcounter{figure}{0}
        \renewcommand{\thefigure}{S\arabic{figure}}%
        \setcounter{equation}{0}
        \renewcommand{\theequation}{S\arabic{equation}}
}

\begin{document}

\title{Phase Diagram and Criticality of the Modified Primitive Electrolyte Model in Bulk and in Inert and Conducting Confinement}
\author{Philipp Stärk}
\affiliation{Stuttgart Center for Simulation Science (SC SimTech),
  University of Stuttgart, 70569 Stuttgart, Germany}
\affiliation{Institute for Computational Physics, University of Stuttgart,
   70569 Stuttgart, Germany}
\author{Alexander Schlaich}
\email{alexander.schlaich@tuhh.de}
\affiliation{
  Institute for Physics of Functional Materials,
  Hamburg University of Technology,
  21073 Hamburg, Germany}

\begin{abstract}
Ionic fluids under conductive confinement are central to technologies such as batteries, supercapacitors, and fuel cells.
Their interfacial behavior governs energy storage and electrochemical processes.
Despite their importance, the thermodynamics of even simple models---such as the charged Lennard-Jones fluid---remain underexplored in this regime.
We present an extended Wang-Landau sampling approach to efficiently compute the density of states of charged mixtures with respect to the particle number.
The method supports simulations in both bulk and confined geometries.
Combined with the Constant Potential Method, it also enables to study effects due to confining electrodes.
We employ this approach to study symmetric, binary mixtures of charged Lennard-Jones particles---the modified Restricted Primitive Model---in bulk, in inert confinement, and in conductive confinement at the potential of zero charge.
Our results show that confinement shifts the vapor-liquid critical point to lower temperatures and higher densities compared to bulk, in line with the classical concept of capillary condensation.
Importantly, conductive boundaries significantly lower the chemical potential of coexistence relative to inert confinement.
These findings offer deeper insight into the phase behavior of ionic fluids in energy-relevant porous environments.
\end{abstract}

\maketitle

\section{Introduction}
\label{sec:motivation}
Ionic liquids and electrolyte solutions are central to applications in energy storage, catalysis, and sensing~\cite{ohno11a}.
Their behavior is dominated by long-range electrostatic correlations, which give rise to rich phase behavior~\cite{orkoulas94a} and challenge mean-field descriptions.
Simplified model fluids provide key insight into these phenomena.
For instance, the Restricted Primitive Model (RPM)---equal-sized charged hard spheres in a dielectric continuum---captures essential collective thermodynamics, including liquid–gas coexistence and criticality~\cite{friedman71a,larsen76a,orkoulas94a,hynninen05a,panagiotopoulos02a,cats21a}.
Lennard-Jones (LJ) fluids with embedded charges extend this framework by incorporating short-range van der Waals interactions~\cite{jones24b}, and are widely used to model ionic systems in empirical molecular dynamics (MD) simulations~\cite{rane14a,anashkin17a,klos20a}.

Confinement, as found in technologically relevant nanopores, is known to drastically alter fluid thermodynamics~\cite{gregg82a, evans86a, charlaixCapillaryCondensationConfined2010, coasneAdsorptionIntrusionFreezing2013}.
Key effects---such as the dielectric behavior of confined electrolytes~\cite{loche20a}, the influence of surface charge~\cite{reinauer24a} on phase equilibria~\cite{gelb97a}, and the relevance of ion correlations~\cite{kondrat10a}---are well documented.
These confinement-induced modifications are precisely the phenomena that control ion adsorption, screening, and charge storage at electrode interfaces.
Consequently, taking these effects into account is essential for developing realistic models of electrode–fluid systems.
Model fluids provide a tractable route to isolate and study these effects in settings relevant to supercapacitors, porous electrodes, and electrochemical devices.
Nevertheless, most existing simulation approaches for such systems (see~\cite{scalfi21a} for a review) rely on the canonical ensemble, neglecting exchange with external particle reservoirs.

To systematically explore ionic thermodynamics in the grand-canonical ensemble, we study a modified version of the Restricted Primitive Model (mRPM): a symmetric 1:1 ionic fluid where particles interact via LJ and Coulomb potentials.
Variants of this model have been considered previously~\cite{rane14a,anashkin17a,klos20a}, but here we adopt the term mRPM to highlight its close conceptual relation to the RPM, while allowing for finite-range repulsion and tunable softness.
To access the grand-canonical thermodynamics of the mRPM, we develop a modified Wang–Landau sampling approach, tailored to charged systems.
Whereas different methods have been employed previously for the study of mixtures in the grand-canonical ensemble like relying on re-weighting \cite{gelb97a}, Gibbs ensemble calculations \cite{orkoulas94a}, or the expanded Wang-Landau approach \cite{desgrangesEvaluationGrandcanonicalPartition2012}, our method enables efficient sampling of ion number fluctuations and grand potential free energy landscapes in both bulk and confined geometries with only minimal modifications to any standard grand-canonical Monte Carlo (GCMC) implemention.
Building on the \textsc{ELECTRODE} package \cite{ahrens-iwers22a} available in \textsc{LAMMPS} \cite{thompson22a}, it further allows for the optional inclusion of fluid-electrode interactions, making it suitable for studying ion adsorption and thermodynamic response in electrochemical environments.

Extensive studies have characterized the thermodynamic properties of LJ fluids in bulk and confinement, including phase equilibria, capillary condensation, and mixture behavior~\cite{panagiotopoulos87a,vanleeuwen91a,desgranges14a,shen05a,faller03a,mejia05a,li12a,evans86a,wang09a,gopalsamy17a,owen15a,tenwolde98a}.
However, even in the bulk the thermodynamic behavior of charged LJ systems remains incompletely understood since long-range electrostatic interactions need to be taken into account.
Previous studies of the charged LJ systems have primarily focused on lattice-based models~\cite{rane14a} or strongly coupled regimes relevant to molten salts~\cite{anashkin17a}.
In contrast, we focus in this study on more moderate Coulomb coupling, representative of aqueous electrolytes, and present a precise characterization of its bulk phase behavior.

Importantly, models of charged fluids under confinement are even less explored.
A notable exception is a recent grand-canonical Monte Carlo study of a charged Weeks–Chandler–Andersen fluid near metallic boundaries using the ICC method~\cite{reinauer24a}.
To our knowledge, we here present the first systematic application of Wang--Landau sampling to a charged LJ fluid---the mRPM---under confinement.
Our approach enables direct access to the Landau free energy of the system and can be extended to arbitrary mixtures and classical electrochemical environments.

\section{Methods}
\label{sec:method}

\subsection{Modified Wang-Landau Sampling for Mixtures}
\begin{figure}[ht]
    \centering
    \def\svgwidth{\columnwidth}
    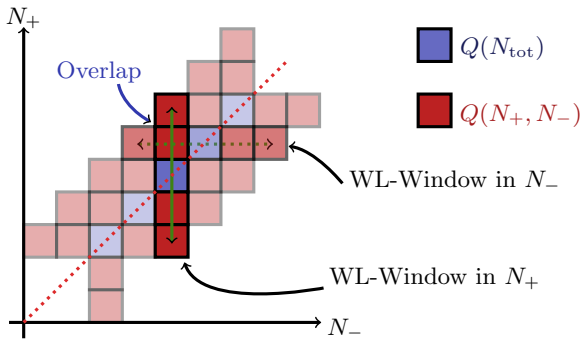

    \caption{Schematic illustration of the sampling scheme proposed in this work. Using small WL sampling runs, which vary the number of one species while keeping the other fixed, we cover the two-dimensional phase space around the neutrality line (dashed red line). Offsets between windows can then be determined from the off-diagonals (red squares), which overlap between (multiple) windows. As an example, we highlight a sampling window in $N_+$ in bold (solid green arrow) and one in $N_-$ (dashed green arrow) with the overlap of this two being labeled. Using the overlap, the offsets between all small sampling patches can be determined. The final result of this sampling are the density of states $Q(N _{\text{tot}})$ along the diagonal, shown as blue squares.}
    \label{fig:criss_cross_illustration}
\end{figure}
To model the thermodynamics of charged fluids, we propose a modified sampling scheme based on the Wang--Landau (WL) method~\cite{wang01a,zhou05a}.
In such systems, key challenges include the long-ranged nature of Coulomb interactions and the necessity to sample binary (or higher-order) mixtures of oppositely charged species while preserving charge neutrality.
Our approach addresses these difficulties by enabling efficient sampling across particle-number space while maintaining electrostatic consistency through the use of the Particle-Particle-Particle-Mesh (P3M)~\cite{hockney88a} method.

The original WL method is a Monte Carlo algorithm designed to estimate the density of states $Q(x)$ along a chosen, discrete reaction coordinate $x$, such as energy or particle number~\cite{wang01a}.
It extends the conventional Metropolis--Hastings framework~\cite{metropolis53a} by iteratively modifying the acceptance probabilities to converge towards uniform sampling in $x$, as explained in detail below.
The bias per discrete value of $x$ that enables traversal of otherwise rarely sampled regions of phase space directly yields the density of state $Q(x)$.

We extend one-dimensional Wang–Landau sampling in $x$ to arbitrary mixtures $(x_1, x_2, \dots)$.
We demonstate its application to a system characterized by two reaction coordinates: the number of positively charged particles, $N_+$, and the number of negatively charged particles, $N_-$.
Sampling is performed independently along these coordinates in separate simulations, and the results are subsequently combined during analysis to reconstruct the two-dimensional density of states $Q(N_+, N_-)$, yielding a two-dimensional free energy surface, while allowing for trivial parallelization.
Due to the symmetry of the system, equilibrium configurations are expected to be charge neutral.
Therefore, we restrict our final analysis to the charge-neutral subspace defined by the diagonal $N_+ = N_-$ in the $(N_+, N_-)$ phase space in a second step and define the total number of fluid particles as $N_{\text{tot}} := N_+ + N_-$.
Evaluating the density of states along this diagonal gives $Q(N_{\text{tot}})$, which defines a one-dimensional reaction coordinate based on the total particle number in neutral systems.

We perform one-dimensional Wang–Landau (WL) sampling runs---described further in the next paragraph---for the total particle number $ N_{\text{tot}}$, varying either the number of positive or negative particles by up to five particles from the neutral configuration, as illustrated in \cref{fig:criss_cross_illustration}.
That is, we sample in the interval $ N_+ \in [N_{\text{tot}} - 5, N_{\text{tot}} + 5] $ while keeping $ N_- $ fixed, or vice versa.
In general, Wang-Landau sampling yields $Q(N_+, N_-)$ up to a constant offset~\cite{wang01a}.
To reconstruct the full two-dimensional density of states, we determine this constant offsets for each window by numerically minimizing the relative distance of overlapping states (see \cref{fig:criss_cross_illustration}), utilizing optimization algorithms of the scipy package~\cite{virtanen20a}.
The details of this procedure are given in the supporting information.
Although combining overlapping WL windows was already proposed by \citeauthor{wang01a}~\cite{wang01a}, our key innovation is to efficiently sample only the narrow region of the two-dimensional space that is of interest for charged mixtures.

Sampling only the density of states for one species, while keeping all others fixed within this window, we drop the subscript +/- for clarity.
In GCMC sampling, the acceptance probability for a move $N \to N'$ is proportional to the Boltzmann factor, $e^{-\beta(E_{N'} - E_N)}$, so that states are sampled according to their equilibrium probability, which can lead to poor sampling in regions where the density of states $Q(N)$ is small.
Wang-Landau sampling addresses this by modifying the acceptance probability to achieve approximately uniform sampling in particle number.
To this end, the Boltzmann weights are multiplied by the ratio of the particle number DOS before and after the trial move,
\begin{equation}
\label{eq:boltzmann_weight}
P_\mathrm{acc}(N \to N') \sim \min\left[1, \frac{Q(N)}{Q(N')} e^{-\beta(E_{N'} - E_N)}\right].
\end{equation}
Here, $\beta = 1/(k_{\text{B}} T)$ is the inverse thermal energy, with $k_{\text{B}}$ being Boltzmann's constant and $T$ the temperature and $E_N$ the energy of the system with $N$ particles.
All states are initialized as $Q(N)=1$, yielding standard GCMC.
Upon each visit to state $N$ this is updated via $Q(N) \to Q(N) \times f_i$, with a refinement factor $f_i$, set to $f_0 = e^4$ in the first iteration.
In each iteration, sampling continues until every histogram bin has been visited at least $ 500 / \sqrt{\ln f_i} $ times, yielding an accuracy $\Delta Q \propto \ln f_i$~\cite{zhou05a}.
After that, the modification factor is updated via the recursive rule $ f_{i+1} = \sqrt{f_i} = f_0^{1/2^{i+1}} $.
This process is repeated until a maximum iteration $ i_{\text{max}} = 16 $ is reached.
Each WL step consists of one insertion or deletion move, followed by five particle displacement moves to ensure decorrelation between samples~\cite{zhou05a}.
At the boundaries of each sampling window, we apply the update scheme described by \citeauthor{schulz03a} to avoid edge artifacts~\cite{schulz03a}.

The end result of the procedure is the density of states of neutral systems by considering the diagonal $N_{\text{tot}}=N_+ + N_-$, from which we determine the Grand potential (Landau free energy) as a function of particle count $N _{\text{tot}}$ by realizing that it approximates the partiation sum \cite{shen05a,errington98a},
\begin{equation}
    \label{eq:landau_free}
\Omega(N _{\text{tot}}) = -k _{\text{B}}T \log[Q(N _{\text{tot}})] - \mu N _{\text{tot}},
\end{equation}
where $\mu$ is the chemical potential of the positive/negative ion (individually).
For further analysis, the free energy function $\Omega(N)$ is fitted with seventh-order polynomials using the method of least squares, with the arbitrary reference choice $\Omega(0) = 0$, unless stated otherwise.
The fit functions then allow for analytical estimation of the location of free energy minima.

In principle, the extended Wang-Landau sampling procedure sketched above also provides data for the density of states for non-neutral systems.
However, we restrict our discussion in this work to neutral systems since non-neutrality poses additional challenges in the interpretation of grand-canonical simulations~\cite{barr12a}.
We do note that analyzing possible, small deviations of these finite systems from charge neutrality might be an interesting avenue for future studies.

In conclusion, the Landau free energy obtained from Wang–Landau sampling, \cref{eq:landau_free}, can be extrapolated to any chemical potential within the sampling in $N_{\text{tot}}$.
This approach was implemented by us with minimal conceptual modifications to the GCMC routines in LAMMPS based on the 22Dec2022 release, and we provide the corresponding modified code on DaRUS~\cite{darus}.
Since it can be trivially parallelized across multiple small sampling windows, it is computationally efficient and scalable to large systems.

\subsection{Modified Restricted Primitive Model: the Symmetric Charged Lennard-Jones Fluid}

In this study, we investigate the critical behavior of particles with pair-wise Lennard-Jones potentials and a unit point charge coincident with the particle's position.
Thus, all particle interactions are described by the following pair-potential $U (R_{ij})$ where particle $i$ and $j$ with valency $z_i, z_j$ have positions $\vec{R}_i$ and $\vec{R}_j$
\begin{equation}
    \label{eq:lj_siunit}
U (R_{ij}) = 4 \ljepsilon \left[ \left(\frac{\sigma}{R_{ij}}  \right)^{12} - \left(\frac{\sigma}{R_{ij}}  \right)^{6}\right] + \frac{e^2 z_i z_j}{4\pi\epsilon_0\epsilon _{\text{r}} R_{ij}},
\end{equation}
and where we use $R_{ij} := \abs{\vec{R}_j - \vec{R}_i}$ to refer to the distance between particle $i$ and $j$ in the units of our simulations.
The parameter $\ljepsilon$ is the characteristic energy of the LJ potential and $\sigma$ can be thought of as the effective diameter of the particles; $\epsilon _{\text{r}}$ is the relative permittivity of the background medium, such as an implicit solvent.
Following \citeauthor{rane14a} we re-write this by introducing reduced units and symmetric charges $z$, yielding~\cite{rane14a}
\begin{equation}
    \label{eq:lj_reduced}
u(r_{ij}) =  4 \left[ \left(\frac{1}{r_{ij}}  \right)^{12} - \left(\frac{1}{r_{ij}}  \right)^{6}\right] + \frac{\alpha}{r_{ij}},
\end{equation}
with
\begin{equation}
    \label{eq:alpha_def}
\alpha := \frac{e^2 z^2}{4\pi\ljepsilon\epsilon_0\epsilon _{\text{r}} \sigma} 
\end{equation}
and $r_{ij} := R_{ij}/\sigma$, $u(r) := U(r)/\epsilon$.
We limit to the case of monovalent ions, $z=\pm1$ and choose $\epsilon _{\text{r}} = 92.636 $, as well as $\sigma = \SI{3}{\angstrom}$, $\ljepsilon = \SI{0.5}{\kilo\joule\per\mole}$, such that $\alpha = 10$.
These values are chosen to be reasonably close to values for typical ions in an implicit aqueous solvent (could be, e.g., a simplified model for NaCl in water), as is evident by comparison to existing classical force fields \cite{loche21a}.
The short-ranged Lennard-Jones interactions are truncated at $r_\text{cut} = 5\sigma$ with an energy shift to remove the discontinuity in the potential; this cutoff closely reproduces full long-range behavior in bulk systems~\cite{schlaich19a}.
The choice of $\alpha = 10$ furthermore allows for comparison with the lattice-based simulations of \citeauthor{rane14a}~\cite{rane14a}.
For simulations of bulk systems, we used periodic boundary conditions of a cubic unit cell with side lengths $L = 8, 9, 10\;\sigma$, in order to perform finite-size scaling analysis.
For confined systems, the next section introduces the relevant definitions.

\subsection{Confined Systems using Hard Walls and the Constant Potential Method}
\begin{figure}[ht]
    \centering
    \def\svgwidth{1\columnwidth}
    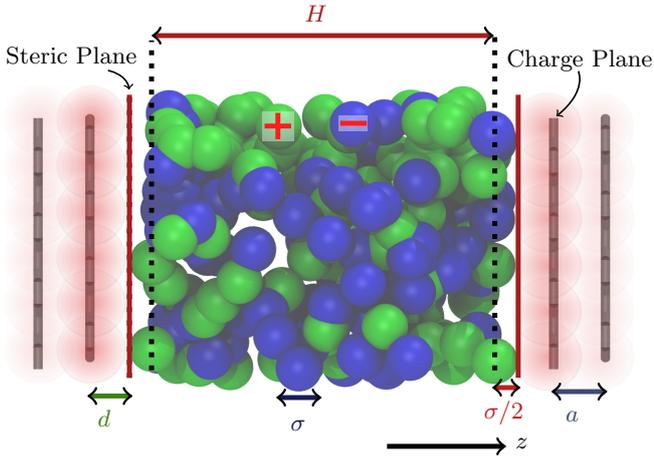

    \caption{
        Schematic illustration of the model slit pore investigated in this work.
        Lennard-Jones particles of characteristic size $\sigma$ are confined between infinitely repulsive steric walls (red lines).
        Their positions are thus restricted to an accessible pore width region $H$ in the $z$-direction (indicated by dashed black lines).
        For systems of polarizable boundaries, Gaussian charge layers (thick gray lines where the Gaussians are insicated as transparent red circles) with a interlayer-separation $a$ are placed at a distance $d = 0.75 \; \sigma$ from the hard walls.
        Black dashes inside the electrodes indicate the underlying graphitic lattice structure.
        Positive and negative charged fluid particles are shown in green and blue, respectively.
    }
    \label{fig:system_illustration}
\end{figure}

Planar confinement was introduced via steric particle-wall interactions using two infinitely repulsive walls separated by a distance $H_\mathrm{p}=H+\sigma$ in the non-periodic $z$ direction (\cref{fig:system_illustration}).
The accessible pore width for the particle centers is $H$ since they cannot approach closer to the hard wall as their radius $\sigma/2$.
In our simulations, we used systems with a accessible pore width of $H/\sigma = 1.67, 3.67, 4.67, 5.67, 6.67, 7.67, 9.67$.
Since the walls are purely repulsive without any material-specific interactions, we refer to this system as \enquote{inert} confinement in the following.

For simulations with polarizable confining walls, we utilize the Constant Potential Method (CPM), implemented in the ELECTRODE package of LAMMPS~\cite{thompson22a}, see Ref.~\cite{ahrens-iwers22b} for details.
In short, the method, originally due to \citeauthor{siepmann95a}~\cite{siepmann95a}, uses Gaussian charges---which are assumed to be centered on the atoms positions---to fulfill a constant potential boundary condition between two electrodes by optimizing the respective amplitudes.
Since the intention of this approach is to mimick conducting electrodes, we will refer to this setup as \enquote{conductive} confinement in the following.

For numerical stability of the CPM in combination with trial insertion moves, the Gaussian charge distributions representing the electrode must be placed at a certain distance outside the effective wall positions.
We choose a distance $d = 0.75\,\sigma$, which offers convinving numerical stability but in general must be considered as specific material paremeter for the solid/electrolyte interface.
The electrode was chosen to represent a graphite-like grid of two layers with a lattice constant $\SI{1.408}{\angstrom}$ and a inter-layer distance of $a = \SI{3.35}{\angstrom}$, see \cref{fig:system_illustration}.
The lateral dimensions were chosen such that the periodic unit cell comprised of 8 by 4 graphite unit cells, resulting in the lateral area $A = \SI{19.68}{\angstrom} \times \SI{17.04}{\angstrom} $ (larger than $5\;\sigma$ in each lateral direction).
The Gaussian charges employed in the CPM, were set to have an inverse width $\eta = \SI{5}{\per\angstrom}$ (following the notation convention of \citeauthor{ahrens-iwers22b}\cite{ahrens-iwers22b}).
To ensure comparability between inert and polarizable confinement, the same steric walls as described above were used in conjunction with the CPM electrodes and the graphitic atoms do not have any interaction with the fluid except the electrostatic ones.

For the planar systems, we use 3d periodic boundary conditions in conjunction with corrections of \citeauthor{yeh99a} to effectively simulate a 2d-periodic system, extending the simulation domain by a factor of three in the non-periodic $z$-direction~\cite{yeh99a}.
Electrostatic interactions are handled via the P3M method~\cite{hockney88a,deserno98b}.
Because error estimates for the Coulomb solver depend on the number of particles~\cite{deserno98a}, which in turn will effect the choice of P3M parameters, one has to be careful to ensure comparability across the different systems and WL windows.
To this end, we tune the P3M parameters only once using the error estimate of \citeauthor{deserno98a} \cite{deserno98a} to an accuracy of $\SI{1e-4}{}$ for the second-largest pore including charges on electrodes and the maximum fluid particle number considered (details see supporting information).
These parameters were then kept fixed for all simulations.

\section{Results and Discussion}
\subsection{Coexistence in Bulk}
\begin{figure}[h]
    \centering
    \includegraphics[width=\linewidth]{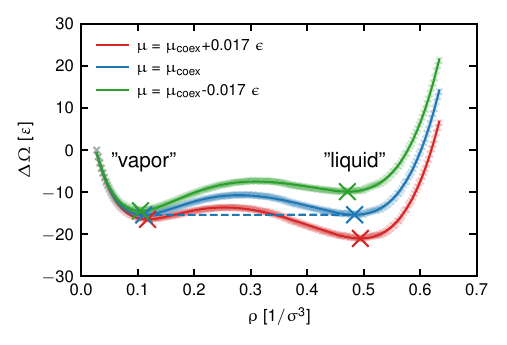}
    \caption{
        Illustration of the Landau free energy landscape as determined from Wang-Landau simulations at $T = 1.55\, \ljepsilon / k _{\text{B}}$ at $L = 9\;\sigma$.
        The free energy landscape belonging to the coexistence chemical potential $\mu _{\text{coex}} = -15.866 \ljepsilon$ is shown in blue, exemplary metastable states red and green.
        Data from WL sampling are shown shown as crosses, lines are seventh-order polynomial fits to the data.
    }
    \label{fig:exemplary_landau_free}
\end{figure}
We in this section first explore the critical behavior of the considered symmetric, charged Lennard-Jones fluid in bulk, before we turn to confinement effects in the next section.
In \cref{fig:exemplary_landau_free}, we provide an exemplary Landau free energy profile as a function of the density $\rho$ for a system at reduced temperature $Tk _{\text{B}}/\epsilon = 1.55$.
The characteristic double-well form of the free energy function points toward the system exhibiting coexistence.
Depending on the chemical potential, the fluid- (red line) or the vapor-phase (green line) are the global minimum, while the other phase represents a metastable minimum (minima are shown by the cross markers).
Through variation of the chemical potential, the chemical potential of coexistence can be determined.
This is defined by the chemical potential $\mu _{\text{coex}}$ at which the minima representing the fluid and gas phases equal, $\Omega(\rho _{\text{l}}) = \Omega(\rho _{\text{v}})$ (see dashed blue line).
The corresponding free energy profile For this particular system at $\mu _{\text{coex}}/\epsilon = -15.866$ is shown as blue line in \cref{fig:exemplary_landau_free}.

\begin{figure}[h]
    \centering
    \includegraphics[width=\linewidth]{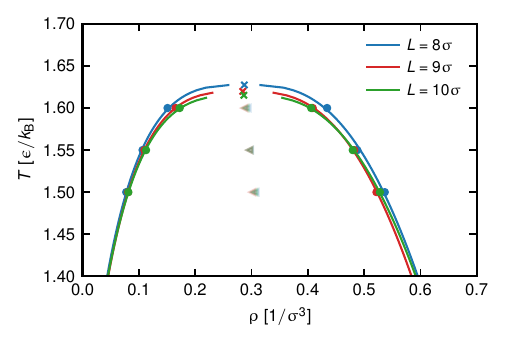}
    \caption{
        Finite size analysis for the vapor-liquid coexistence of the mRPM with $\alpha=10$.
        The coexistence densities determined from the extended Wang-Landau sampling scheme for are shown as round symbols, the extrapolated critical point $(\rho _{\text{c}}, T _{\text{c}})$ as a cross and the arithmetic mean of the coexisting phases is given by transparent triangles.
        The color code represents the simulation box size $L$.
    }
    \label{fig:rho_vs_T}
\end{figure}

Following this procedure for different systems, we obtain $\mu _{\text{coex}}, \rho _{\text{v}}, \rho _{\text{l}}$ at given $T$ and simulation box size $L$.
The coexistence densities $\rho _{\text{v}}$ and $\rho _{\text{l}}$ are shown as circles \cref{fig:rho_vs_T}.
The critical behavior of charged Lennard-Jones fluids is expected to belong to the Ising universality class, which we also investigate more below~\cite{rane14a}.
This allows us to extrapolate the critical point according to~\cite{schlaich19a,smit92a}
\begin{equation}
    \label{eq:density_scaling}
    (\rho _{\text{l}} - \rho _{\text{v}}) = A(T - T _{\text{c}}) ^{\beta _{\text{i}}},
\end{equation}
where $\beta _{\text{i}} = 0.326419 $ is the second Ising exponent in three dimensions and $A$ a fitting parameter.
The same critical scaling holds true for the arithmetic mean, i.e.,
\begin{equation}
    \label{eq:arithmetic_scaling}
    \left(\frac{\rho _{\text{l}} + \rho _{\text{v}}}{2} \right) = B (T - T _{\text{c}})^{\beta _{\text{i}}},
\end{equation}
which allows us to extrapolate to the critical density $\rho _{\text{c}}$ for $T \to T _{\text{c}}$, with $B$ being another fitting parameter.
Results for both fits are shown in the supporting information, Fig.~S3.
The mean densities (lhs of \cref{eq:arithmetic_scaling}) are shown as transparent triangles in \cref{fig:rho_vs_T} together with the estimated coexistence lines determined from both fits as well as the critical density $\rho _{\text{c}} (L)$ and temperature $T_{\text{c}} (L)$ for each system size (cross markers).

Rigorous finite-size analysis \cite{ferrenberg91a,wilding95b} allows to extrapolate our results to the infinite-box limit via a linear fit of $T_{\text{c}}(L)$ versus $L^{-(1 + \Theta)/\nu_{\text{i}}}$, using Ising exponents $\nu_{\text{i}} \approx 0.6289$ and $\Theta \approx 0.54$, shown in the supporting information, Fig.~S3.
This yields an estimate for the bulk critical temperature of $T_{\text{c}} = \SI{1.598 \pm 0.001}{}\, \ljepsilon / {k_\mathrm{B}}$, with the uncertainty determined from the covariance of the linear fit.
Using on-lattice simulations \citeauthor{rane14a} reported a critical temperature of $T_{\text{c}} = \SI{1.6891 \pm 0.0004}{}\, \ljepsilon / {k_\mathrm{B}}$~\cite{rane14a} for $\alpha=10$, which is significantly higher than our estimate.  
This is, however, expected since our truncated and shifted LJ model differs strictly from their work using cutoff of $r_\mathrm{cut} = 100\sigma$; larger cutoffs are known to increase the critical temperature~\cite{schlaich19a}.
Given this fact we conclude that our results are well in line with the literature and thus validating our modified WL sampling approach.

In order to further evaluate whether the assumption of Ising universality is well-founded, we also investigate the scaling of the surface tension close to the critical temperature.
For this purpose, we created a liquid-vapor interface by elongating the simulation box of length $L = 10\sigma$ in $z$-direction by a factor of three.
The surface tension $\gamma$ is calculated using the mechanical definition as~\cite{kirkwood49a,rowlinson13a}
\begin{equation}
    \label{eq:surface_tension}
\gamma = \frac{L_z}{2} \expval{ P_{zz} - \frac{P_{xx} + P_{yy}}{2}},
\end{equation}
where the ensemble average is given by $\expval{\cdot}$ and $P_{\alpha\alpha}$ is the $\alpha,\alpha$-components of the pressure tensor.
For this purpose, we used additional NVT simulations carried out with LAMMPS with an integration timestep of $1\,\text{fs}$, sampling the pressure tensor every $\SI{1}{\pico\second}$ for a total amount of $\SI{1e6}{}$ samples after an equilibration time of $\SI{100}{\pico\second}$.
As the interface gets more diffuse the closer one gets to the critical point, we limit the analysis to temperatures $T\,k_\mathrm{B} / \ljepsilon \in \{1.425, 1.45, \dots, 1.5\}$, see Fig.~S2 of the supporting information.
Fitting \cref{eq:surface_tension} to our simulation data, we estimate $\nu \approx \SI{0.622 \pm 0.021}{}$, which is shown in Fig.~S1 and which is in very good agreement with literature values for Ising critical scaling~\cite{leguillou80a}.
This supports the conclusion that the restricted Primitive Model falls into the Ising universality class.

\subsection{Criticality in (Conducting) Confinement}
\begin{figure}
    \centering
    \includegraphics[width=\linewidth]{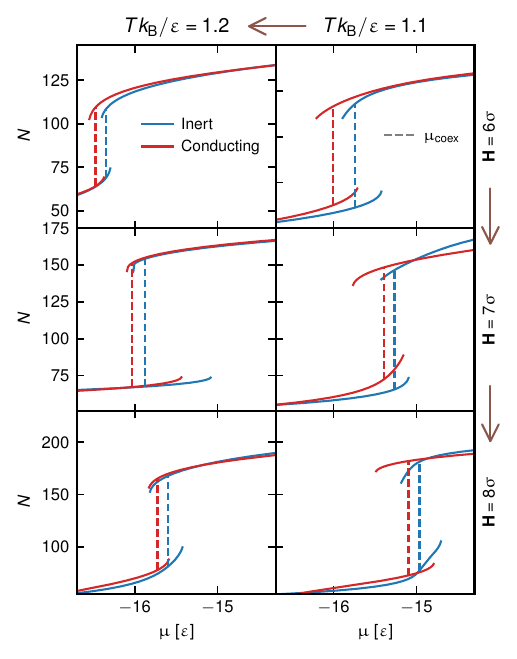}
    \caption{Adsorption isotherms for three pore widths and two different temperatures. We compare conducting and electrostatically inert confinement for every system. Shown are the vapor/liquid particle counts $N _{\text{v/l}}$ vs. the chemical potential. The chemical potential of coexistence $\mu _{\text{coex}}$ in each system is indicated by the dashed line. The direction of increase of control variables between panels is shown as brown arrows. Clearly, we see a shift towards higher values in $\mu _{\text{coex}}$ for a decrease in temperature and a increase in pore size. A consistent shift towards higher $\mu _{\text{coex}}$ can also be observed when going from conducting to inert confinement.
    }
    \label{fig:isotherm}
\end{figure}

We now turn to the behavior of the mRPM in confinement.
In \cref{fig:isotherm}, we show adsorption isotherms for various pore sizes and temperatures.
Each isotherm denotes the particle number corresponding to the minima in the free energy landscape at a given chemical potential, cf.\ \cref{fig:exemplary_landau_free}.
At low $\mu$, only a stable vapor phase is observed, while increasing $\mu$ allows the emergence of a metastable liquid-like phase.
Dashed lines indicate the positions of liquid–vapor coexistence, with metastable states appearing on either side.
Increasing the chemical potential in the reservoir at these sub-critical temperatures thus leads to capillary condensation.
At high chemical potentials, the fluid phases approach a saturation density.
Upon adsorption (increasing $\mu$), the system eventually follows the liquid branch of the isotherm, while desorption (decreasing $\mu$) follows the vapor branch, leading to emergence of a hysteresis loop.

For a given pore size and temperature, comparison between the electrostatic boundary conditions (inert vs.\ conducting) reveals a significant influence of wall polarizability on the condensation behavior of the confined fluid.
In the presence of polarizable walls, the condensation occurs at a lower chemical potential compared to inert confinement.
Given that this shift can only be due to the difference in wall-fluid interactions, all else being equal, this suggests that image charge effects lower the chemical potential of coexistence.
Incresing the temperature shifts the coexistence chemical potential $\mu _{\text{coex}}$ to lower values, in line with the decrease of the coexistence pressure with increasing temperature in bulk systems.
Likewise, increasing the pore size shifts $\mu _{\text{coex}}$ to higher values, as the system is expected to approach bulk-like behavior in the limit $H \to \infty$.

To further elucidate on the influence of wall polarizability we show in the supporting information, in Fig.~S4 of the Supplementary Information we exemplary show density profiles for the vapor phases at the same chemical potential.
Importantly, strong layering of the fluid at the interface is observed for both conducting and inert confinement, as expected for fluids near hard walls~\cite{sullivanStructureSimpleFluid1978}.
However, a significantly more pronounced density peak is observed in the case of conducting confinement, hinting at stronger adsorption due to image charge effects and in line with the increased adsorbed amount in the vapor phase observed in \cref{fig:isotherm} for the conducting interfaces.
Yet, when comparing both systems at their respective coexistence chemical potential (Fig.~S2 in the Supplementary Information), the density profiles agree nearly perfectly, in line with the fact that the adsorbed amount is very similar for both conducting and inert confinement just before/after capillary condensation.
This indicates that the strongly adsorbed fluid layer is rather independent of $\mu$ at the interface.

Differences between both systems shown in \cref{fig:isotherm} are also more pronounced for $T = 1.1 \, \ljepsilon / k_{\text{B}}$, whereas both systems seem more similar closer to the critical temperature for $T = 1.2 \ljepsilon / k_{\text{B}}$.
This behavior is also expected, as image charge effects are anticipated to be more significant further away from the critical point, since the fluid structure at the interface competes with entropy.
These differences underscore the significant impact of the electrode properties on adsorption characteristics and highlight the potential for tuning phase behavior through conductivuty changes or applied potentials when porous materials are in contact with external reservoirs.

\begin{figure*}
    \centering
    \includegraphics[width=\linewidth]{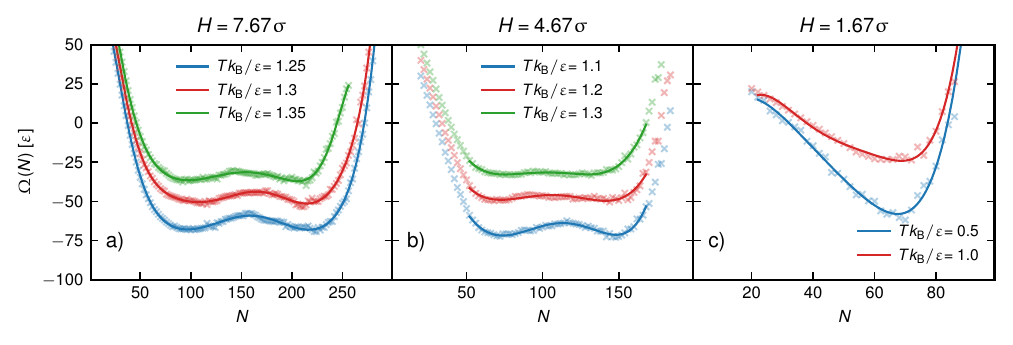}
    \includegraphics[width=.32\linewidth]{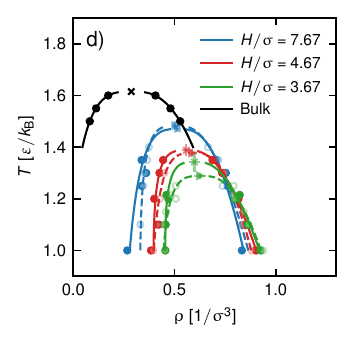}
    \includegraphics[width=.32\linewidth]{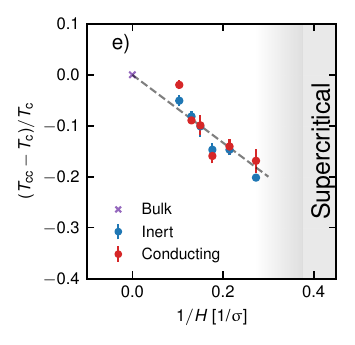}
    \includegraphics[width=.32\linewidth]{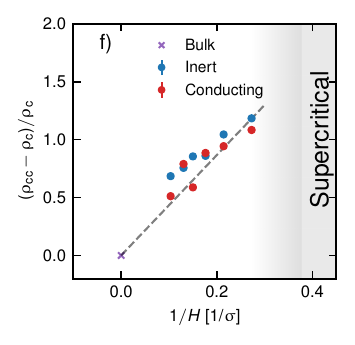}
    \caption{
        (a-c): Landau free energy landscapes for three exemplary conducting pores at different temperatures.
        Profiles are shifted by $10\ljepsilon$ for clarity.
        Raw data is shown as crosses as well as polynomial fits as lines.
        For (a) + (b), the chemical potential corresponds to coexistence.
        In (c) the fluid is always super critical independent of the temperature within the range considered, as evident from the single minimum present.
        (d): Exemplary coexistence lines in confinement and in bulk.
        Full circle markers show the location of vapor/liquid minima at coexistence, the crosses shows the location of the critical points.
        Dashed lines show the Ising fit to the data in inert confinement.
        (f-e): Relative change in capillary critical temperature $T_{\text{cc}}$ and density $\rho_{\text{cc}}$ with respect to the bulk critical temperature/density vs.\ inverse pore size.
        The pore sizes for which no critical behavior could be observed are indicated by the grey shaded region.
        }
    \label{fig:criticality_ov}
\end{figure*}

In \cref{fig:criticality_ov} (a) - (c), we show exemplary free energy landscapes for different temperatures and pore sizes for the case of conducting confinement.
At relatively large pore sizes, $H = 7.67 \sigma$ and $ 4.67 \sigma$ one can clearly observe two distinct minima in analogy to the bulk case above.
Surprisingly, we do not observe such two such minima for the smallest pore studied, $H/\sigma =  1.67 $.
Rather, the presence of a minimum in the Landau free energy---even at the temperatures as low as $T = 0.5 \, \ljepsilon / k_{\text{B}}$---indicates that the fluid is always supercritical.
The smallest pore size for which criticality was observed in our simulations is $ H = 3.67\sigma $.
This suggests the existence of a critical pore size $ H_{\text{c}} $, with $ 1.67\sigma \leq H_{\text{c}} < 3.67\sigma $, below which criticality does not occur.
This is in contrast with the mean-field model of \citeauthor{evans86a}, as well as with literature on the uncharged LJ fluid in confinement~\cite{panagiotopoulos87a,vishnyakov01a,evans86a}, where critical behavior is found also for such small pore sizes.
Given the mean-fields model's neglect of correlation effects and the absence of long-ranged interactions in these studies, we speculate that the explicit inclusion of correlation effects due to electrostatic interactions in our model causes this the difference.
This is also in line with expectations from the physics of ionic fluids, where strong ionic correlations can lead to pronounced structuring effects, emergence of in-plane organization and even the formation of ionic crystals at sufficiently high coupling strengths~\cite{levinElectrostaticCorrelationsPlasma2002,najiElectrostaticInteractionsStrongly2005}.

In \cref{fig:criticality_ov}~(d), we show the phase diagrams for confined fluids between conducting (solid symbols) and insulating (open symbols) walls of different reduced pore widths \( H \), together with the bulk coexistence curve (corresponding to the data for \( L = 10\sigma \) in \cref{fig:rho_vs_T}).
The strong density layering near the walls leads to a significant increase in the averaged density both in the gas and in the liquid phase within the pore compared to the bulk coexistence curve [black data in \cref{fig:criticality_ov}~(d)].
Yet, the coexistence densities well follow the Ising scaling laws, \cref{eq:density_scaling,eq:arithmetic_scaling}, as shown by the fits to the to the corresponding data (solid and dashed lines).
The narrower shape of the coexistence envelopes indicates that density differences between $\rho _{\text{v}}$ and $\rho _{\text{l}}$ become smaller, in line with the high density layer of ions near the surface present in both gas and liquid phases.

Confinement leads to a strong reduction of the critical temperature and shifts the critical density toward higher values as the pore width decreases, indicated by the extrapolated critical points in \cref{fig:criticality_ov}~(d).
The relative shifts of capillary critical temperatures $T _{\text{cc}}$ and densities $\rho _{\text{cc}}$ with respect to bulk are shown as a function of the inverse pore size $1/H$ in \cref{fig:criticality_ov}~(e) and (f), respectively.
The critical pore size $H_\text{c}$, for which only supercritical fluids were observed, is indicated by the shaded gray region.
Based on mean-field assumptions, \citeauthor{evans86a} found, for a hard-sphere fluid in confinement with exponentially decaying attractive potential and in the limit of $H/\sigma \gg 1$~\cite{evans86a}
\begin{equation}
    \label{eq:evans_criticality}
\frac{T _{\text{cc}} - T _{\text{c}}}{T _{\text{c}}} \propto \frac{1}{\lambda H},
\end{equation}
where $\lambda$ represents the characteristic length of the attractive fluid-wall potential and $T _{\text{c}}$ is the critical temperature of the fluid in bulk.
This relation is in line with scaling arguments of \citeauthor{nakanishi82a}~\cite{nakanishi82a} and was found to be applicable for simple confined fluids both in experiments and simulations~\cite{thommesPoreCondensationCriticalPoint1994, morishigeAdsorptionHysteresisPore1998, morishigeCapillaryCondensationNitrogen2002, coasneAdsorptionIntrusionFreezing2013, schlaich19a}.
As shown in \cref{fig:criticality_ov}~(e), such a linear relation between the capillary critical temperature and inverse pore size also seems to hold for the mRPM.
Noteworthy, we find a similar behavior also for $\rho_\text{cc}$ to hold, see \cref{fig:criticality_ov}~(f).
The linear behavior is followed by all systems, even for remarkably small pore sizes close to $H _{\text{c}}$.
Surprisingly, little difference is observable for the $T _{\text{cc}}$ between conducting and inert confinement.
However, $\rho _{\text{cc}}$ are slightly enhanced for the inert case compared to conducting confinement, in line with the shift to higher coexistence chemical potentials (\cref{fig:isotherm}).

Summarizing, the main effect of conducting vs.\ inert confinement on the thermodynamic behavior appears to be a shift in the coexistence chemical potential (i.e.\ the pore filling pressure), with more pronounced differences in the adsorption isotherms appearing near the critical chemical potential $\mu_{\text{coex}}$ and for lower temperatures, as noted above.
These findings are also consistent with previous studies~\cite{breitsprecher15a}, which reported only minor differences in the density profiles, when comparing the adsorption profiles of ions near polarizable versus inert confinement.

\section{Conclusion}
In this work, we demonstrated how an extended Wang-Landau sampling method can efficiently be utilized to sample mixtures, specifically binary ionic fluids in the Grand canonical ensemble.
While the approach is general, we paid particular focus to describing the vapor-liquid transition.
We presented precise estimates for the critical point of a symmetric charged Lennard-Jones fluid in bulk with rigorous finite size scaling.
We further substantiated the classification of this mRPM fluid within the Ising universality class via investigation of surface tensions near the critical point.

Despite modest differences in the structure of the liquid at a given chemical potential, we found that conductivity strongly alters adsorption isotherms and leads to a significant reduction in the critical chemical potential.
Surprisingly, our results provide evidence for a critical pore size $H _{\text{c}}$, below which the fluid only exhibit supercritical behavior and which presumably appears due to strong ion correlations.
This insight might be practically relevant for the design of functional material, such as phase change materials or for the utilization of pores in aiding chemical reactions, which are known to sometimes be positively enhanced by supercriticality~\cite{poliakoff15a}.

We have examined the confinement-induced shift of the critical point of the in planar confinement, with both conducting and electrostatically inert pore walls.
As pore size decreases, the critical temperature was observed to decrease together with a concurrent increase in the critical density.
The effects of conductive or inert nature of the confinement were found to be small, although the critical density is slight decrease for the conducting system at given pore size.
This might be rationalized by image charge effects that primarily influence the first adsorbed layers.

Importantly, while we have applied the extended Wang–Landau sampling to a relatively simple model system, the approach itself is flexibile and broadly applicable to more complex mixtures or realistic molecular systems, including arbitrary confining geometries.
This generality and numerically efficiency through trivial parallelizability makes the method a promising tool for future studies aiming to probe phase behavior under experimentally relevant conditions.
As the mRPM captures the essential physics of typical ionic fluids, we expect wide qualitative applicability of our findings.
However, several limitations remain.
For instance, the symmetric, monovalent Lennard-Jones fluid considered here does not capture the complexity of real ionic fluids used in applications.
Introducing asymmetry and valency differences could offer promising directions for future study.
Similarly, the model used for the confining electrode could be refined further and especially the effect of applied potentials might be of interest for future research.


\section*{Supplementary Material}

\section*{Acknowledgments}
Funded by Deutsche Forschungsgemeinschaft (DFG, German Research Foundation) under Germany's Excellence Strategy - EXC 2075 – 390740016 and SFB 1333/2 – 358283783.
We acknowledge the support by the Stuttgart Center for Simulation Science (SimTech).
We further acknowledge funding from the Deutsche Forschungsgemeinschaft (DFG, German Research Foundation) through the Compute Cluster grant no. 492175459 and 261833929.
We thank Alexander Reinauer and Svyatoslaw Kondrat for fruitful discussions and Christian Holm for support in this project.

\section*{Author declarations}

\subsection*{Conflict of Interest}

The authors have no conflicts of interest to declare.

\subsection*{Author Contributions}

\textbf{Philipp Stärk}: Conceptualization (supporting); Investigation (equal); Methodology (equal); Software (lead); Validation (lead); Writing – original draft (lead); Writing – review \& editing (equal).
\textbf{Alexander Schlaich}: Conceptualization (lead); Investigation (equal); Funding acquisition (lead); Methodology (equal); Project Administration (lead); Supervision (lead); Writing – review \& editing (equal).

\section*{Data Availability Statement}
All raw data, including all density of state histograms for all systems and the modified LAMMPS source code is available on DaRUS, the data repository of the University of Stuttgart \cite{darus}.

\bibliography{Library.bib}

\onecolumngrid
\clearpage

\renewcommand{\thefigure}{S\arabic{figure}}
\renewcommand{\theequation}{S\arabic{equation}}
\setcounter{figure}{0}
\setcounter{equation}{0}

\beginsupplement
\section{Additional Figures}
\begin{figure}[h]
    \centering
    \includegraphics[width=.5\linewidth]{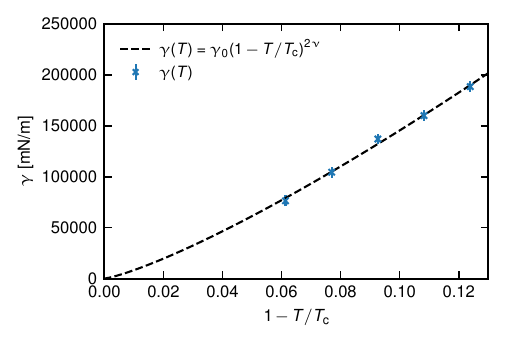}
    \caption{
      Surface tension scaling close to the critical temperature.
      Blue markers show simulation results.
      The dashed line denotes the fit of the expected scaling behavior of the Ising universality class to the simulation data, resulting in $\nu = \SI{0.622 \pm 0.021}{}$ well in line with the established literature $\SI{0.62997097\pm0.00000012}{}$ \cite{changBootstrapping3dIsing2025}.
      }
    \label{fig:gamma_scaling}
\end{figure}

\begin{figure*}[h]
    \centering
    \includegraphics[width=0.49\linewidth]{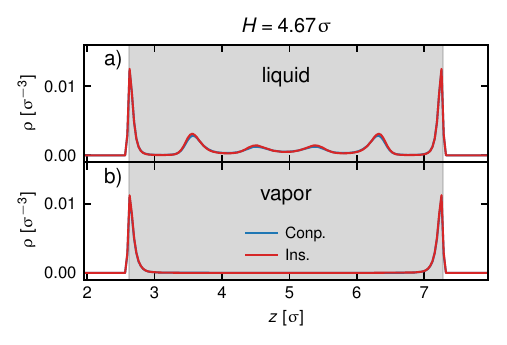}
    \includegraphics[width=0.49\linewidth]{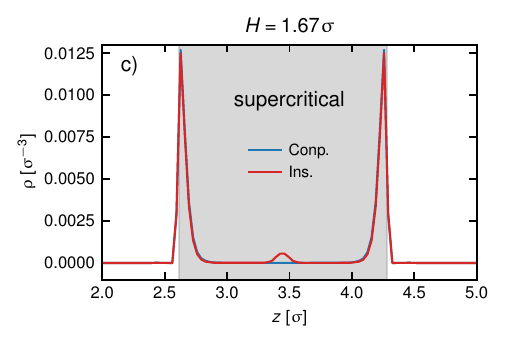}
    \caption{
      Density profiles of exemplary systems. In panel a) and b) we show results for a pore with a electrode distance $H = 4.67\sigma$ at the coexistence chemical potential for the temperature $T k _{\text{B}}/\epsilon = 1.0$.
      In panel c) we show results for a pore with electrode distance $H = 1.67\sigma$. 
      The regions accessible to the fluid particles are shown in gray.
      For the larger pore, the system exhibits coexistence and we show density profiles for the liquid and vapor at the point of coexistence.
      The smaller system only exhibits super critical behavior and we show an exemplary density profile density profile for 27 ion pairs in the pore.
      In all panels, red lines were derived from simulations with conducting electrodes and blue for simulations with non-conducting walls.}
    \label{fig:dens_profs}
\end{figure*}

\section{Details on the Combination of the independently sampled WL Windows}
As explained in the main text, the Wang-Landau simulations are run for individual windows sampling a sub-space of the desired reaction corrdinate (the number of positive and negative particles in our case).
The individual windows describe the density of states for the interval $[N-5, N+5]$ in the deviation of positive or negative particles from the line of neutrality (see also Figure 1 in the main text).
Note that due to symmetry, only windows for positive deviations in particle number need to be sampled explicitly, reducing the numerical effort by a factor two.

The data from individual windows then need to be processed to obtain a continuous density of states over the entire range of particle numbers, meaning that consistent offsets between windows corresponding to an arbitrary constant when converting the density of states to a Landau free energy must be determined.
For numerical efficiency, we do not optimize offsets for all windows at once, but subgroup the data into overlapping blocks of size 8, i.e.\ 8 windows for the positive and 8 windows for the negative species being considered at once.
For these 16 windows the offsets are optimized simultaneously.
To provide a concrete example: the first \enquote{block} consists of windows where $N_+$ is varied in the interval $N_+ \in [0, 10]$, the window $ N_+ \in [1, 11]$ and so on until $N_+ \in [8, 18]$ together with the windows $N_- \in [0, 10] \dots$ $N_- \in [8, 18]$.

For each of these blocks, we then determine offsets between overlapping windows as follows.
First, we select the histograms with identical particle numbers $(N_+, N_-)$ in all windows of a block.
Then, a first guess for the offsets is obtained by minimizing the $L_1$ norm between overlapping data points by minimizing the loss function
\begin{equation}
    \mathcal{L} = \frac{1}{N_{\text{overlap}}} \sum_{i,j} \left| Q_1(N_{ij}) + o_1^{(i)} - Q_2(N_{ij}) - o_2^{(j)} \right|,
\end{equation}
where $o_1^{(i)}$ and $o_2^{(j)}$ are the offset parameters for windows $i$ and $j$ in the two datasets and automatic differentiation with just-in-time compilation is used for numerical efficiency.
Last, a least-squares optimization refines the same objective, and we choose the solution yielding the lower residual.

This procedure is repeated until constructing blocks with an overlap of $n_{\text{overlap}}=3$ common states on the line of neutrality until the full particle number range is covered.
After determining intra-block offsets as described, blocks are concatenated by minimizing the $L_2$ norm between the last $n_{\text{overlap}}$ points of block $k$ and the first $n_{\text{overlap}}$ points of block $k+1$:
\begin{equation}
    \Delta o_k = \arg\min_{\delta} \left\| \ln\Omega_k^{(\text{end})} - (\ln\Omega_{k+1}^{(\text{start})} + \delta) \right\|_2
\end{equation}
Cumulative offsets $o_k = \sum_{i=1}^k \Delta o_i$ then align all blocks into a continuous profile.

\section{Ewald Summation Parameters for Simulations}
The parameters for the P3M method were determined once for a large system, at $H = 7.67 \sigma$ in order to keep consistency between systems.
To this end, the maximum number of particles was placed in the system and a charge of $0.02 e$ placed on the electrode atoms, allowing us to let the LAMMPS P3M error estimate tune the parameters of the method to achieve the desired relative force accuracy of $10^{-4}$.
The large system with the maximum number of charges and inclusion of charges on the electrode was chosen since the error estimates for the P3M method depend on the total number of charges present in the system.
This allows for a consistent comparison between all free energy profiles.
The resulting P3M parameters of this procedure were a $k$-grid of 8, 8, 10 in $x, y, z$-respectively (with $z$ the axis along the pore) and the smearing parameter of the Ewald sum as $g _{\text{Ewald}} = 0.13424533 $.
These parameters were kept fixed for all confined systems.
Following the analogue procedure for bulk systems, we set the Ewald $k$-grid to 15, 15, 15 and the $g _{\text{Ewald}}$ parameter to 0.2145291 for all bulk systems.

\section{Finite Size Scaling}
\begin{figure}[h]
    \centering

    \begin{tikzpicture}
        \node[anchor=south west, inner sep=0] (img) {\includegraphics[width=0.49\linewidth]{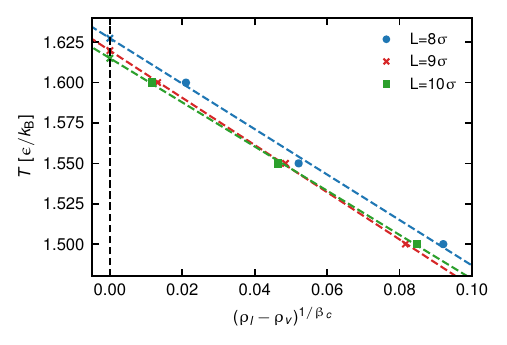}};
        \node[anchor=north west, font=\bfseries] at (img.north west) {(a)};
    \end{tikzpicture}
    \hfill
    \begin{tikzpicture}
        \node[anchor=south west, inner sep=0] (img) {\includegraphics[width=0.49\linewidth]{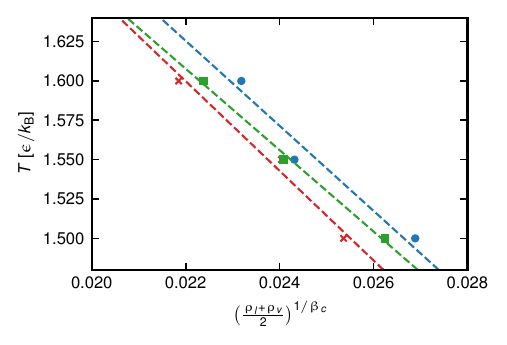}};
        \node[anchor=north west, font=\bfseries] at (img.north west) {(b)};
    \end{tikzpicture}

    \begin{tikzpicture}
        \node[anchor=south west, inner sep=0] (img) {\includegraphics[width=0.49\linewidth]{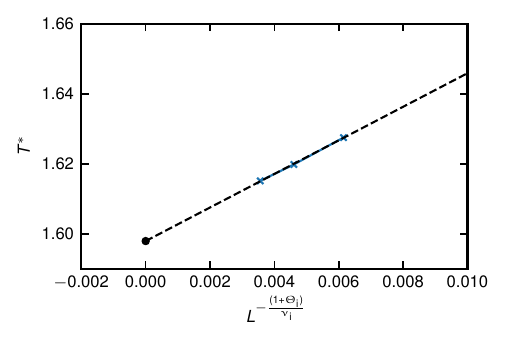}};
        \node[anchor=north west, font=\bfseries] at (img.north west) {(c)};
    \end{tikzpicture}

    \caption{Scaling plots of difference between the vapor and liquid densities, their arithmetic mean, as well as finite size scaling fit to extrapolate to the critical temperature in the infinite box size limit.}
    \label{fig:ffs_temp}
\end{figure}

As was already described in the main text, we performed finite size scaling for the bulk system.
To this end, we used three systems with cubic side-length $L/\sigma = 8, 9, 10$.
Three temperatures were investigated for each system, namely $T k _{\text{B}}/\epsilon = 1.5, 1.55, 1.6$ and the liquid and vapor densities determined from the Landau free energy.
Next, we fit \cref{eq:density_scaling} of the main text to the liquid and vapor densities, as shown in \cref{fig:ffs_temp}~(a).
Extrapolating these curves towards vanishing differences between vapor and liquid densities we determine the critical temperature $T _{\text{c}}$ for each of the systems.
\Cref{fig:ffs_temp}~(b) shows the corresponding fit of \cref{eq:arithmetic_scaling} in the main text to the mean densities.
Extrapolating this scaling relation to the $T _{\text{c}}$ that follows from \cref{fig:ffs_temp}~(a), we retrieve the critical density $\rho _{\text{c}}$, which is then shown in \cref{fig:rho_vs_T} in the main text.
As also described there, in order to extrapolate the critical temperature to the infinite system limit, we follow~\cite{ferrenberg91a,wilding95b} and fit $T _{\text{c}}$ vs. $L^{-\frac{(1+\Theta _{\text{i}})}{\nu _{\text{i}}}}$, which is shown in \cref{fig:ffs_temp}~(c).

\section{Density Profile for a Fixed Chemical Potential}
\begin{figure}[h]
    \centering
    \includegraphics[width=0.49\linewidth]{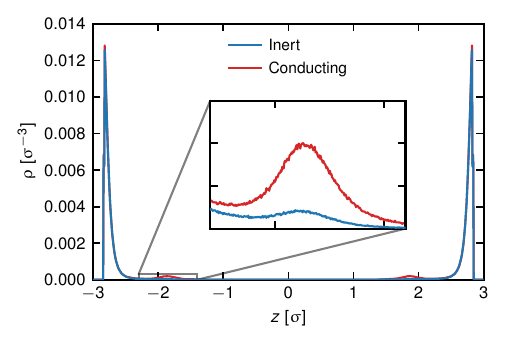}
    \caption{Density profiles for the equilibrium vapor densities at $\mu = -15.48 \epsilon$ in a system with/without conducting confinement at pore size $H = 5.67 \sigma$. The influence of conductivity is clearly visible in the increased density close to the wall and significantly decreased density in the second layer as shown in the inset.}
    \label{fig:density_vapor_phase}
\end{figure}

In order to investigate the differences between both inert and conducting confinement, we performed NVT simulations with the equilibrium vapor density for the exemplary system with $H/\sigma = 6.67$ at $T^* = 1.1$ and $\mu = -15.48 \epsilon$.
The density profile of this system is shown in \cref{fig:density_vapor_phase}.
This effectively compares inert and polarizable confinement at a fixed chemical potential.
Differences between both systems are only minor with both systems exhibiting strong wetting of the surface.
Profiles show a faint second layer, which is further highlighted in the inset.
The main difference between conducting and inert confinement seems to be the increase in adsorbed ions near the surface for the conducting/polarizable wall.
However, the similarity between both systems is likely the reason why they differ only slightly in their thermodynamic behavior, as was already discussed in the main text.

\end{document}

%% file: figures/criss_cross_illustration.pdf_tex
\begingroup%
  \makeatletter%
  \providecommand\color[2][]{%
    \errmessage{(Inkscape) Color is used for the text in Inkscape, but the package 'color.sty' is not loaded}%
    \renewcommand\color[2][]{}%
  }%
  \providecommand\transparent[1]{%
    \errmessage{(Inkscape) Transparency is used (non-zero) for the text in Inkscape, but the package 'transparent.sty' is not loaded}%
    \renewcommand\transparent[1]{}%
  }%
  \providecommand\rotatebox[2]{#2}%
  \newcommand*\fsize{\dimexpr\f@size pt\relax}%
  \newcommand*\lineheight[1]{\fontsize{\fsize}{#1\fsize}\selectfont}%
  \ifx\svgwidth\undefined%
    \setlength{\unitlength}{566.63501169bp}%
    \ifx\svgscale\undefined%
      \relax%
    \else%
      \setlength{\unitlength}{\unitlength * \real{\svgscale}}%
    \fi%
  \else%
    \setlength{\unitlength}{\svgwidth}%
  \fi%
  \global\let\svgwidth\undefined%
  \global\let\svgscale\undefined%
  \makeatother%
  \begin{picture}(1,0.51230508)%
    \lineheight{1}%
    \setlength\tabcolsep{0pt}%
    \put(0,0){\includegraphics[width=\unitlength,page=1]{criss_cross_illustration.pdf}}%
    \put(0.02295027,0.48418513){\color[rgb]{0,0,0}\makebox(0,0)[lt]{\lineheight{1.25}\smash{\begin{tabular}[t]{l}$N_+$\end{tabular}}}}%
    \put(0.51213805,0.00703528){\color[rgb]{0,0,0}\makebox(0,0)[lt]{\lineheight{1.25}\smash{\begin{tabular}[t]{l}$N_-$\end{tabular}}}}%
    \put(0,0){\includegraphics[width=\unitlength,page=2]{criss_cross_illustration.pdf}}%
    \put(0.71414333,0.4371513){\color[rgb]{0.09411765,0.10980392,0.34117647}\makebox(0,0)[lt]{\lineheight{1.25}\smash{\begin{tabular}[t]{l}$Q(N _{\text{tot}})$\end{tabular}}}}%
    \put(0.71382502,0.33514014){\color[rgb]{0.64705882,0.11372549,0.11764706}\makebox(0,0)[lt]{\lineheight{1.25}\smash{\begin{tabular}[t]{l}$Q(N_+, N_-)$\end{tabular}}}}%
    \put(0,0){\includegraphics[width=\unitlength,page=3]{criss_cross_illustration.pdf}}%
    \put(0.51486995,0.08474725){\color[rgb]{0,0,0}\makebox(0,0)[lt]{\lineheight{1.25}\smash{\begin{tabular}[t]{l}WL-Window in $N_+$\end{tabular}}}}%
    \put(0.54529528,0.23302364){\color[rgb]{0,0,0}\makebox(0,0)[lt]{\lineheight{1.25}\smash{\begin{tabular}[t]{l}WL-Window in $N_-$\end{tabular}}}}%
    \put(0,0){\includegraphics[width=\unitlength,page=4]{criss_cross_illustration.pdf}}%
    \put(0.10014391,0.40107118){\color[rgb]{0.19215686,0.22352941,0.68235294}\makebox(0,0)[lt]{\lineheight{1.25}\smash{\begin{tabular}[t]{l}Overlap\end{tabular}}}}%
    \put(0,0){\includegraphics[width=\unitlength,page=5]{criss_cross_illustration.pdf}}%
  \end{picture}%
\endgroup%

%% file: figures/illustration_system.pdf_tex
\begingroup%
  \makeatletter%
  \providecommand\color[2][]{%
    \errmessage{(Inkscape) Color is used for the text in Inkscape, but the package 'color.sty' is not loaded}%
    \renewcommand\color[2][]{}%
  }%
  \providecommand\transparent[1]{%
    \errmessage{(Inkscape) Transparency is used (non-zero) for the text in Inkscape, but the package 'transparent.sty' is not loaded}%
    \renewcommand\transparent[1]{}%
  }%
  \providecommand\rotatebox[2]{#2}%
  \newcommand*\fsize{\dimexpr\f@size pt\relax}%
  \newcommand*\lineheight[1]{\fontsize{\fsize}{#1\fsize}\selectfont}%
  \ifx\svgwidth\undefined%
    \setlength{\unitlength}{357.53183921bp}%
    \ifx\svgscale\undefined%
      \relax%
    \else%
      \setlength{\unitlength}{\unitlength * \real{\svgscale}}%
    \fi%
  \else%
    \setlength{\unitlength}{\svgwidth}%
  \fi%
  \global\let\svgwidth\undefined%
  \global\let\svgscale\undefined%
  \makeatother%
  \begin{picture}(1,0.69831499)%
    \lineheight{1}%
    \setlength\tabcolsep{0pt}%
    \put(0,0){\includegraphics[width=\unitlength,page=1]{illustration_system.pdf}}%
    \put(0.79251137,0.02796231){\color[rgb]{0,0,0}\makebox(0,0)[lt]{\lineheight{1.25}\smash{\begin{tabular}[t]{l}$z$\end{tabular}}}}%
    \put(0,0){\includegraphics[width=\unitlength,page=2]{illustration_system.pdf}}%
    \put(0.15654609,0.06188612){\color[rgb]{0.24313725,0.50980392,0.0745098}\makebox(0,0)[lt]{\lineheight{1.25}\smash{\begin{tabular}[t]{l}$d$\end{tabular}}}}%
    \put(0,0){\includegraphics[width=\unitlength,page=3]{illustration_system.pdf}}%
    \put(0.7790922,0.61432089){\color[rgb]{0,0,0}\makebox(0,0)[lt]{\lineheight{1.25}\smash{\begin{tabular}[t]{l}Charge Plane\end{tabular}}}}%
    \put(0,0){\includegraphics[width=\unitlength,page=4]{illustration_system.pdf}}%
    \put(0.01724291,0.61873744){\color[rgb]{0,0,0}\makebox(0,0)[lt]{\lineheight{1.25}\smash{\begin{tabular}[t]{l}Steric Plane\end{tabular}}}}%
    \put(0,0){\includegraphics[width=\unitlength,page=5]{illustration_system.pdf}}%
    \put(0.86900161,0.0691049){\color[rgb]{0.23921569,0.28627451,0.43921569}\makebox(0,0)[lt]{\lineheight{1.25}\smash{\begin{tabular}[t]{l}$a$\end{tabular}}}}%
    \put(0,0){\includegraphics[width=\unitlength,page=6]{illustration_system.pdf}}%
    \put(0.744419,0.07414361){\color[rgb]{0.7372549,0.12941176,0.13333333}\makebox(0,0)[lt]{\lineheight{1.25}\smash{\begin{tabular}[t]{l}$\sigma/2$\end{tabular}}}}%
    \put(0.47203769,0.68237731){\color[rgb]{0.64705882,0.11372549,0.11764706}\makebox(0,0)[lt]{\lineheight{1.25}\smash{\begin{tabular}[t]{l}$H$\end{tabular}}}}%
    \put(0,0){\includegraphics[width=\unitlength,page=7]{illustration_system.pdf}}%
    \put(0.45053855,0.05590872){\color[rgb]{0.09411765,0.10980392,0.34117647}\makebox(0,0)[lt]{\lineheight{1.25}\smash{\begin{tabular}[t]{l}$\sigma$\end{tabular}}}}%
    \put(0,0){\includegraphics[width=\unitlength,page=8]{illustration_system.pdf}}%
  \end{picture}%
\endgroup%